# Provenance of the Cross Sign of 806 in the Anglo-Saxon Chronicle: A possible Lunar Halo over Continental Europe?

Yuta Uchikawa (1-2)*, Les Cowley (3), Hisashi Hayakawa (4-5)**, David M. Willis (5-6), and F. Richard Stephenson (7)

(1) University of Tokyo, 7-3-1 Hongo, Bunkyō, Tokyo 1138654, Japan

(2) Fitzwilliam Museum, University of Cambridge, Trumpington St, Cambridge CB2 1RB, UK

(3) Atmospheric Optics, www.atoptics.co.uk, Norfolk, UK

(4) Graduate School of Letters, Osaka University, Toyonaka, Osaka, 5600043, Japan (JSPS Research Fellow)

(5) Science and Technology Facilities Council, Rutherford Appleton Laboratory, Harwell Campus, Didcot, OX11 0QX, UK

(6) Centre for Fusion, Space and Astrophysics, Department of Physics, University of Warwick, Coventry, CV4 7AL, UK

(7) Department of Physics, University of Durham, Durham, DH1 3LE, UK

* yu210@cam.ac.uk; uchikawayuta@g.ecc.u-tokyo.ac.jp

** hisashi.hayakawa@stfc.ac.uk; hayakawa@kwasan.kyoto-u.ac.jp

## Abstract

While graphical records of astronomical/meteorological events before telescopic observations are of particular interest, they have frequently undergone multiple copying and may have been modified from the original. Here, we analyze a graphical record of the cross-sign of 806 CE in the Anglo-Saxon Chronicle, which has been considered one of the earliest datable halo drawings in British records, whereas another cross-sign in 776 CE has been associated with the aurora. However, philological studies have revealed the later 806 event is derived from Continental annals. Here, records and drawings for the 806 event have been philologically traced back to mid-9th Century Continental manuscripts and the probable observational site identified as the area of Sens in northern France. The possible lunar halos at that time have been comprehensively examined by numerical ray tracing. Combined with calculations of twilight sky brightness, they identify a visibility window supporting monastic observation. Cruciform halos are shown to be fainter and rarer than brighter and more commonplace lunar halos. Physically credible cloud ice crystal variations can reproduce all the manuscript renditions. The manuscript records prove less than desirable detail but what is presented is fully consistent with a lunar halo interpretation. Finally, the possible societal impacts of such celestial events have been mentioned in the context of contemporary coins in Anglo-Saxon England and the Carolingian Empire. These analyses show that we need to trace their provenance back as far as possible, to best reconstruct the original event, even if graphical records are available for given astronomical/meteorological events.

## 1) Introduction:

Historical records from before the era of instrumental observations have occasionally provided valuable information for space science. This is the case with variable solar activity (*e.g.*, Vaquero, 2007; Vaquero and Vazquez, 2009; Schlegel and Schlegel, 2011; Usoskin, 2017; Hayakawa et al., 2019c), supernovae (*e.g.*, Stephenson and Green, 2002; Smith, 2013), comets (*e.g.*, Stephenson *et al.*, 1985; Kronk, 1999; Hayakawa *et al.*, 2017b), and earth rotation (*e.g.*, Newton, 1972; Stephenson *et al.*, 2016; Soma and Tanikawa, 2016; Gonzalez, 2018). Likewise, in this context, several historical meteorological records have been interpreted as atmospheric optics events (Boyer, 1987; Tape and Moilanen, 2006). Among them are a Russian record of atmospheric optics at Astrakhan in 1670 CE (Usoskin *et al.*, 2017), graphical evidence such as Vädersolstavlan drawing in 1535 CE (Cowley, 2009) and Matthew Paris's 1123 halo account (Lewis, 1987), medieval chronicle records (Newton, 1972), multiple Suns in the Roman chronicles and coins (Woods, 2012), and Emperor Constantine's vision (Weiss, 2003).

Accordingly, the International Astronomical Union has recently called for urgent need of preservation of historical records including these pre-modern ones (Pevtsov *et al.*, 2019). Upon consulting these historical records, the graphical records of astronomical events (Figure 1) are of particular importance, as they frequently convey further details over the verbal records (*e.g.*, Stephenson and Willis, 1999; Willis and Stephenson, 2001; Hayakawa *et al.*, 2017a, 2017b). However, these graphical records have frequently undergone multiple copying and may have been modified from the original, as also proved to be the case in some oriental sunspot drawings (Hayakawa *et al.*, 2018, 2019b; Fujiyama *et al.*, 2019).

Figure 1: The earliest known graphical records of sunspots (left) and candidate aurorae (right). The earliest known sunspot record is considered to be the one in the Chronicle of Worcester (CCC MS 157, f. 192 v, courtesy of 2019 Corpus Christi College, Oxford) and dated on 1128 CE December 8 (Stephenson and Willis, 1999; Willis and Stephenson, 2001). The earliest known candidate auroral record based on actual observations is considered to be the one in the Chronicle of Zūqnīn (MS Vat.Sir.162, f. 150v, courtesy of 2019 Biblioteca Apostolica Vaticana) and dated in 771/772 CE

(Hayakawa *et al.*, 2017b). Both of these drawings are considered to be autograph drawings (Willis and Stephenson, 2001; Hayakawa *et al.*, 2017b).

In this context, the *Anglo-Saxon Chronicle* (ASC) (Swanton, 2000), the primary historical source for Anglo-Saxon England (c. 450-1066) represented by seven manuscripts stemmed and developed independently from a lost original (MSS A to G; written in Old English except for the Old English/Latin bilingual MS F), contains reports of astronomical and meteorological phenomena such as eclipses, comets, and aurorae during the medieval period (*e.g.*, Newton, 1972; Egler, 2002; Beard, 2005; Schlegel and Schlegel, 2011; Härke, 2012). Among such phenomena, two celestial signs 'of the cross' were reported; one in 776 CE and the other in 806 CE. The first 'cross' in 776 CE has been recently subjected to consideration in the context of the extreme solar particle event in 774/775 CE recorded via cosmogenic-isotope proxies in tree rings and ice cores (*e.g.*, Miyake et al., 2012; Usoskin *et al.*, 2013; Mekhaldi *et al.*, 2015; Büntgen *et al.*, 2018; Uusitalo *et al.*, 2018). The first cross-sign, namely the 'red sign of Christ' in 776 CE has been considered a plausible auroral record after detailed philological and scientific analyses (Schlegel and Schlegel, 2011, p. 43; Usoskin *et al.*, 2013; Stephenson, 2015). While we need to be slightly cautious about possible contamination of atmospheric optics (*e.g.*, Usoskin *et al.*, 2017; Carrasco *et al.*, 2017), comparison with early modern visual observations showed that this report seems plausibly an auroral display (Hayakawa *et al.*, 2019a). This discussion suggested, together with other contemporary records (Usoskin *et al.*, 2013; Hayakawa *et al.*, 2017; Stephenson *et al.*, 2019) that the solar event of 774/775 CE occurred near the maximum phase of the solar cycle.

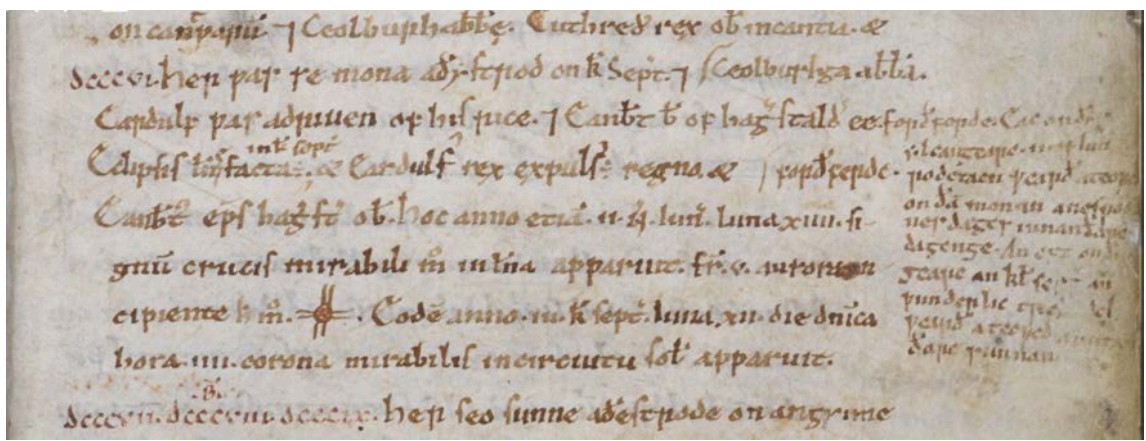

Figure 2: The 'cross' sign in the lunar event on 806 CE June 4 in the Anglo-Saxon Chronicle MS F (British Library MS Cotton Domitian A viii, f. 51r), adapted from MS with courtesy of the British Library Board, adapted from Hayakawa et al. (2019a).

On the other hand, the account of the second 'cross' sign on 806 CE June 4 with its accompanying drawing has been described as a lunar halo (Figure 2; Newton, 1972; Neuhäuser and Neuhäuser, 2015). This report is found only in Manuscript F, a bilingual version of the ASC in Old English and Latin, written in Canterbury in the early twelfth century (Baker, 2000, p. 59; Swanton, 2000, p. 59). Newton (1972) cited this as a meteorological report recorded in texts in the British Isles, as well as in France and Germany, while he has not clarified their textual relationship. Nevertheless, philological analyses show that this report and drawing are probably copied from earlier continental annals (Baker 2000, pp. xlvi-xlvii, 59; Hayakawa *et al.*, 2019a). Therefore, the present article traces the philological genealogy of the lunar record of 806 CE June 4 as far back as possible, transcribing and translating the various records, reconstructing the observation, and considering the actual provenance of the 806 report. At the same time, this record still shares some important insights into the history of Anglo-Saxon England and contemporary Europe, together with the celestial sign in 776 CE, in their terminology of 'the sign of the cross' and their apparent relationship with images on contemporary coins. Therefore, we revisit the provenance of the 806 drawings, scrutinise the lunar halo hypothesis, and return to 'the sign of the cross' and its possible historical impact on the society of Anglo-Saxon England.

## 2) Record of the 806 'cross' in ASC and continental annals: Philological Analysis

The celestial record on 806 CE June 4 is one of multiple celestial spectacles around 806, including solar and lunar eclipses as well as solar and lunar haloes not only in the ASC but also in the continental annals (Swanton, 2000, p. 59; *Chronicon in Aetatis Sex Divisum*, in Migne, 1852, col. 131; see also Frobesius, 1739, p. 17). This record the Latin Version of the ASC MS F reads: "Also in this year, on the 4th of June, the 14th day of the Moon, the sign of the cross in a remarkable fashion, appeared at the Moon on Thursday at dawn, like this #", accompanied with a probable solar halo record stating "a wonderful crown … around the Sun" (see Appendix 1). Taken literally, the text and drawing are slightly inconsistent: the text places the cross sign 'at the Moon', whereas the drawing shows the cross sign around the Moon. The text and its literal interpretation would rule out a halo hypothesis since a lunar halo display is "around" or "outside" the Moon. However, this could have resulted from the MS F scribe being unable to find a proper word to describe the location of the cross against the Moon and resorting to the most general term. After that, he might think it helpful to add the drawing to clarify his intention.

Given that the lunar record in 806 CE is found only in MS F without parallel records in any other ASC manuscripts, it is worth considering the provenance of this record (see Hayakawa *et al.*, 2019a). This is because we frequently have citations and hearsays from other source materials in the ASC manuscripts, as exemplified by an annular solar eclipse record in 809 CE, which is also derived from continental annals together with the 806 event. While the recorded solar eclipse on 809 CE July 16 in ASC MS F (Baker, 2000, p. 60) is definitely consistent with an annular solar eclipse on that very day, the darkness then seems to have been exaggerated. For instance, at Trondheim (N63°26′, E10°24′), one of the places where the eclipse would be central, the magnitude has been calculated to be 0.97 at 10.8 LT. Hence only about 94% of the Sun's disc would be covered at maximal phase, which would hardly lead to darkness. The magnitude at Sens (N48°12′, E3°17') has been computed to be only 0.60 at 10.0 LT. An eclipse of such small magnitude would be unlikely to be noticed either here or in central Europe. The path of annularity passed close to the Arctic Circle and crossed central Europe, not France. However, given that some of the celestial records are copied from external sources, it is quite likely that this eclipse record is also based on external sources or hearsays. Indeed, the reports in the Old English version of Orosius' *The Seven Books of History Against the Pagans* written in the 890s attest to commercial communications between Western Europe and Northern Europe; one of these reports relates the navigation of Ohthere from Hålogaland (the northernmost province of Norway at that time) to the White Sea (Godden, 2016, pp. 36-49). This is even applicable to Francia, where similar contacts with Denmark and Sweden were attempted in the form of Christian missionaries in the early 9th century (the most famous among them is St. Ansgar, the apostle of the North). Therefore, as in the 806 halo record, the reports for the solar eclipse are also probably based on hearsays or external sources. Alternatively, while the Old English translation in ASC MS F could mean the actual darkening of the Sun, the original Latin texts of the continental manuscripts, 'solis eclipsis apparuit' only denote 'a solar eclipse appeared', and thus the darkness was not necessarily exaggerated if the partial solar eclipse was observed at Sens in some way, e.g. through cloud, which reduced the glare of the Sun and enabled the solar disc to be seen with the unaided eye.

Concerning the lunar record in 806 CE, it is found only in MS F without parallel records in any other ASC manuscripts. This is due to two facts. First, MSS E and F shared the same lost exemplar (from which the scribes copied the text), which is supposed to have an entry of a lunar eclipse under 806 CE (Baker, 2000, pp. xxix-xxxix; Irvine, 2003, pp. xxxviii-xxxix, xl-xliii); and MS D also mentions an eclipse in the same year, but it is not of the Moon but the Sun, which seems to be a scribal error (Cubbin 1996, p. 19). All three manuscripts share much in common and are ultimately derived from the so-called 'Northern Recension' of the ASC, which shows strong interest in astronomy beyond the 806 event (e.g. *sub anno* (hereafter, *s. a.*) 744, 800, 802; Cubbin, 1996, pp. xviii-xxi; Baker, 2000, p. xxix; Irvine, 2003, pp. xxxvi-lviii), which can be traced back to Alcuin and Bede (Eastwood, 2013, pp. 311--315).

Second, the scribe of MS F also had access to a manuscript based on a now lost Latin chronicle, the *Winchester Chronicle* (WC; Liebermann 1879, pp. 56-83), which in turn incorporated some entries from continental sources[1]. The entry of 806 CE in MS F begins with the usual combination of an Old English text and its Latin translation concerning a lunar eclipse. After that, the scribe inserted the description of the lunar and the solar halo from the lost WC in Latin, as if he found it opportune to add some information on phenomena related to the Sun or the Moon in the same year. These constitute the main text and the Old English translation of the WC entry was inserted into the margin. This explains why this halo record in 806 CE employed a different Old English word "*rodetacn* (cross-sign)" (MS F, f.51r.) from the other records of a "*Cristes Mel* (sign of Christ)" in 776, while they share the same term "sign of the (Lord's) cross (*signum* (*dominicæ*) *crucis / crucis signum*)" in Latin. Thus, it is probable that '*rodetacn*' in 806 was translated from Latin and '*Cristes Mel*' in 776 was translated into Latin.

While the scribe of MS F might have had no idea where his information ultimately came from, the editor of the critical edition of MS F (Baker 2000, pp. xlvi-xlvii, 59) pointed out that the text of the 806 entry resembles the ones in two continental annals called *Annales Sancti Maximini Trevirensis* (ASMT) and *Annales Laudunenses Et Sancti Vincentii Mettensis Breves* (ALESVMB). These manuscripts have been located to MS 2500 in Trier Stadtbibliothek/Archiv (former Koblenz, Görrische Bibliothek 16; Boeck et al., 1990; Pertz, 1841, pp. 5-7) and MS Phill. 1830 in Berlin Staatsbibliothek (Pertz 1888, pp. 1293-1295), respectively. ASMT is listed by Newton (1972), together with ASC MS F and a 12th-century French manuscript[2]. However, further investigation revealed that there are five 9th-century continental manuscripts with this entry: ASMT, ALESVMB, *Annales Sanctae Columbae Senonensis* (ASCS) (Biblioteca Apostolica Vaticana, Regin. Lat. 755; Pertz, 1826, pp. 102-109), *Annales Lemovicenses* (AL) (Bibliothèque nationale de France, Latin 5239; Pertz, 1829, pp. 251-252; Martène and Durand, 1717, cols. 1400-1402), and *Annales Floriacenses* (AF) (Bibliothèque nationale de France, Latin 5543; Pertz, 1829, pp. 254-255; Vidier, 1965, pp. 217-220).

These five annals include the celestial drawing in the margin of a computational table of Easter. They derived from the same original annals from 708 CE to 840 CE, which were assumed to have been composed in Sens or Laon (See below). Judging from the wording and construction of the text, Schröer (1975) supposes more than one person contributed to the production. For our purpose, it should be noted that the person responsible for the astronomical records may be different from the scribe of the main annals (Schröer, 1975, pp. 83-4). Although the original is now lost, the 806 entry is regarded as a contemporary observation because of its exceptional accuracy. The manuscripts listed in the previous paragraph are divided into three groups by Schröer (1975). According to his stemma or genealogical table of the manuscripts, AF, ASCS, ALESVMB, and ASMT are distant from the lost original (X) by two generations and AL by three. However, it is not mentioned which is the closest to the original. In order to determine it, the content of the entry should be scrutinised. Firstly, AL is discarded because it derives from ASCS and shares annals with ASCS until 861 CE when it is supposed to have been copied. AF, which shares annals with ASCS until 953 CE, is also excluded since it omits the description of the solar halo. Hence, the ASCS was compiled at the monastery of St. Columba in Sens sometime between 853 CE and 861 CE and represents one tradition. As of ALESVMB and ASMT, it is known that the earlier part of ASMT until 840 was composed in *c*. 840 CE at Laon, and that of ALESVMB compiled *c*. 875 CE has perfect concordance with it before these two annals were transferred to St. Maximin in Trier in 876 CE and St Vincent of Metz after its foundation in 968 CE respectively (Boeck et al. 1999, p. 25; Pertz 1888, p. 1293). Thus only ASCS and ASMT deserve further consideration. They show marked differences in their wording and word order (Appendix); while ASCS has 'anno incarnationis dominicae' and 'quasi', ASMT lacks them. It is more reasonable to lose some words than to add them during copying. Moreover, ASCS's 'feria 5. prima aurora incipiente, quasi hoc modo +' contrasts with ASMT's 'hoc modo # feria 5. prima aurora incipiente'. Either 'feria 5. prima aurora incipiente' or '(quasi) hoc modo +'must have been skipped and added immediately by the scribe. It is unlikely that such a revered sign of the cross with its drawing escaped the eyes of the monk. Therefore, ASCS seems to retain the original wording.

---

[1] A twelfth-century chronicle founded on WC also has the text and the drawing of the 806 event (London, British Library, Cotton Nero A. viii, f. 25r-25v; Liebermann, 1879, p. 63).
[2] The latter (*Chronica Domni Rainaldi Archidiaconi S. Mauricii Andegavensis*, c. 1152) was composed too late to be regarded as important in our discussion.

As for the provenance of the original (X) or at least of the astronomical reports, Sens is more likely than Laon considering these facts: (1) While annals derived from Y (copied from X at Sens) describe the date of the solar eclipse in 840 CE fully, those from Z (copied from X at Laon) omit some detail (See Figure 3 below); (2) only ASCS and AL reports 'fiery edges' in the sky in 861 CE; and (3) only ASCS continues to record astronomical events in 868 CE (comets), 909 CE (a comet), 919 CE (fiery and bright edges of various colours in the sky), etc. (the 909 event is also recorded in AF, but in 905 CE). There seems to have existed a continuous interest towards astronomical phenomena among the monks in Sens. Also in this respect, we should assume the text of ASCS as the closest reproduction of the original.

On the other hand, analysis of the drawing indicates a more complex history of information transfer. Figure 3 shows the variation of the drawings among these manuscripts. The drawing in ASCS is like a cross *pattée* (a cross with expanding arms) with a dot in its centre. It also has some blurred figure at the bottom of the cross. On the other hand, the other annals show a voided cross with an annulet in the crossing. Every drawing differed from each other in one way or other, but they are more or less the same except the one in ASCS. Thus, although ASCS's text seems to follow the original most faithfully, we should suppose the drawings in the others as its closest reproduction, probably based on the actual observation by the scribe of the original or some contemporary person who gave him this information. Also, AL probably had access to the lost exemplar in Sens (Y in Figure 3) in addition to ASCS.

There is another interesting variation between the cross drawings. Those derived from Sens (ASCS, AF and AL) have serifs at the cross tips. Laon derived crosses do not. There is some stylistic progression from ASCS to AF then AL, suggesting that ASCS could be closest to the original.

The transmission from these continental manuscripts to ASC MS F via a lost Winchester Chronicle is also complicated. While MS F followed ASCS in the wording of the 806 entry with minor differences, its drawing looks more similar to the ones in the other four manuscripts, but has an annulet filled with ink in different colour. Judging from the text and the drawing, the lost WC and eventually ASC MS F might be derived from Y, AF, AL or their descendants. One thing quite interesting to our discussion is that the practice of recording astronomical phenomena in these continental monasteries was probably built upon the example of Bede. Indeed, ASMT and ALESVMB have records of solar eclipses in 538 CE and 540 CE, which were reported by Bede in his *Ecclesiastical History of the English People* (Book V, chapter 24; Colgrave and Mynors, 1969, pp. 562-563). The manuscript of ASMT also contains other works by Bede, *De Temporibus, De Temporum Ratione, and De Natura Rerum*, which deal with scientific knowledge at that time and astronomy is one of his main concerns (Boeck, 1990, pp. 21-24). Therefore, the insertion of the 806 event in ASC MS F is as if it found all the way back to Bede's homeland to join the series of astronomical records kept in the Northern Recension of the ASC, which follows the tradition of Bede.

Therefore, we conclude that the lunar event in 806 CE was observed not in Anglo-Saxon England but in Continental Europe, probably in Sens. Identifying five more annals with these lunar drawings, we have reconstructed their probable genealogy. This genealogy tells us that the original observational report in Sens had been cited multiple times and the lunar drawing had appeared in ASC MS F, gradually and slightly transforming its shape.

Figure 3: Cross-signs in the relevant annals and their reconstructed genealogical tree, where black squares represent possible lost manuscripts between the reproduced manuscript figures: (from left to right) (a) AF: Bibliothèque nationale de France (Latin 5543, f. 16r), (b) ASCS: Biblioteca Apostolica Vaticana (Vat. Reg. Lat. 755, f.11r), (c) AL: Bibliothèque nationale de France (Latin 5239, f. 13r), (d) ASC MS F: London, British Library (MS Cotton Domitian A viii, f.51r), (e) ASMT: Stadtbibliothek/Stadtarchiv Trier (MS 2500, f.6r), and (f) ALESVMB: Staatsbibliothek zu Berlin (MS Phill. 1830, f.7v).

## 3) Lunar Halo Hypothesis

We postulate that the lunar ice halos were observed during dawn on 806 CE June 4 at Sens (N48°12′, E3°17'). The sky conditions and most favourable observation times for naked-eye visibility are established. We examine variations in the cross appearance corresponding to those in different manuscript renditions. As described below, the findings are also valid for Laon (N49°34′, E3°37′) and indeed for latitudes 3° south or north of Sens. Location longitude is unimportant as all times were local solar.

Ice halos are white or prismatic rings of curvilinear shapes in a sunlit or moonlit sky. They are produced by refraction and reflection of light by small hexagonal ice crystals. In temperate regions these populate high cirrus and some altostratus cloud. In colder climates the crystals may be close to the ground as "diamond dust". The crystals are hexagonal plates (P) and hexagonal columns with their long axis nearly horizontal (C) and randomly oriented hexagonal prisms. Halo formation and characteristics are detailed in Tape (1994).

Halos generated by moonlight are no different to solar halos except in one critical aspect – they are faint because the illuminant, even the full Moon, is some 400,000 times fainter than the Sun (Land and Irvine, 1973). Figure 4 shows a photograph of a lunar halo display in a dark sky. The faint cross centred on the Moon was formed by the intersection of the paraselenic circle and lunar pillar. Its naked-eye appearance was likely fainter. Lunar and even solar halo crosses formed by cirrus clouds in temperate climes are rare. The impressive cruciform halos often seen in photographs are from low-level diamond dust in the polar regions or otherwise severe sub-zero temperatures. We reject diamond dust in non-mountainous, middle-latitude Europe in June as not being meteorologically credible. In addition, a weather occurrence needed to create diamond dust, with its likely human and agricultural impact, would most certainly be in many records. We discuss only halos from high level clouds.

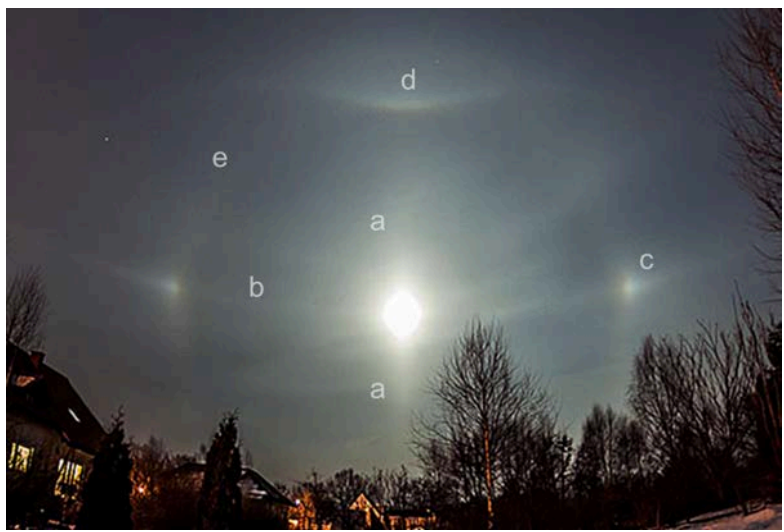

Figure 4: Lunar halo display with cruciform shape. Photographed in Silesia, Poland (N51°11′, E17°17′) by Leszek Sulich on 2013 March 27; A mixture of plate (P), column (C), and randomly oriented (R) habit ice crystals in cirrus clouds formed the display: (a) Lunar pillars from columns C, (b) Paraselenic circle from P and C, (c) Paraselene or moon dog, P, (d) Upper tangent arc C, and (e) 22° halo from R.

Maximum lunar halo visibility is when the Moon is at or near full and high in a dark sky. The near full condition existed in 806 CE with the Moon a computed 13.9 days old, comparing well with the 14 days of the record. The time of observation and the corresponding sky brightness are major issues because halo visibility depends on contrast with the surrounding sky. Additional sky brightness from scattered sunlight at high altitude reduces the contrast. Also, halo brightness and geometry can change with lunar altitude and thus time.

The statement "at the first dawn (*prima aurora incipiente*)" implies an early phase of morning twilight. We investigate the onsets of the three modern divisions of twilight: Astronomical with the Sun 18° below horizon, Nautical 12° and Civil 6°. The onset times and the corresponding Moon altitudes are given in Table 1 for Sens. With the exception of astronomical twilight which is still present at midnight, the times at Laon (N49°34′, E3°37′) differ by no more than a few minutes and the lunar altitudes are within a degree. These differences are unimportant for predictions of halos or sky brightness. Further calculations show that, conservatively, our predictions are valid for latitudes within ±3° of that of Sens. Longitude is unimportant as all times are local solar.

Sky brightness 20° from the Moon (the region of the cross) was calculated using the model of Krisciunas and Schaefer (1991). Its predictions, for a clean gaseous atmosphere, are an approximate best case for lunar halo visibility, as dust, aerosol and the necessary halo forming thin cirrus cloud also scatter light and further reduce halo contrast.

Table 1: Moon altitude and sky brightness 20° from the Moon during dawn at Sens

| Twilight division | Onset time, local solar | Moon altitude, deg | Ratio of total sky brightness* to Moon contribution | Moon contribution to sky brightness, % |
|---|---|---|---|---|
| Astronomical, AT | 00:58 | 21.6 | 1.04 | 96 |
| Nautical, NT | 02:24 | 12.5 | 2.1 | 47 |
| Civil, CT | 03:19 | 5.0 | 9.9 | 11 |
| Sunrise | 04:00 | -0.3 | | |

*Total brightness at 20° azimuth from Moon = background sky brightness + lunar contribution

The ratio of total sky brightness to that from scattered moonlight is near unity at the start of astronomical twilight, *i.e.* almost all background skylight is from the Moon. This is the best condition for halo visibility other than pillars, that we can expect. At nautical twilight onset, already half of sky brightness comes from sunlight scattered at high altitude but potential halo visibility remains good. In contrast, the sky brightness at civil twilight onset is ten times greater than that from moonlight alone. These predictions and observational experience suggest a visibility window for *any* halos extending from pre-dawn to before the onset of civil twilight.

We predict possible halos by numerical simulation. Several million individual light rays are traced through mathematical representations of a combination of ice crystals habits. The HaloSim program of Cowley and Schroeder (1998) was used. The ray tracings automatically showed *all* the halos formed by the chosen crystal population.

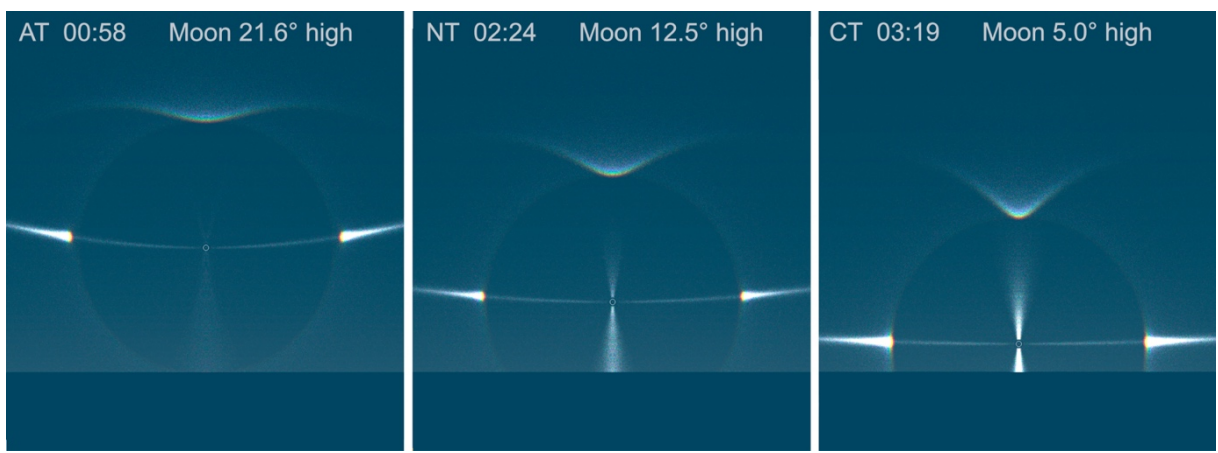

Figure 5: Ray tracing predictions for lunar halos at the onset of each twilight phase at Sens. Dark blue is ground, lighter blue is a generic sky representation. The Moon is the circle at cross centre. Crystal populations: Horizontal columns 1° dispersion c/a=2 20%, random columns c/a=2 8%, plates 8° disp. basal faces function only 15%, plates 1° disp. c/a=0.1 49%, plates 3° disp. c/a=0.1 8%. 8 million incident rays per tracing.

Figure 5 shows predictions for the three twilight onsets. A cross is evident at all the lunar altitudes sampled although at the start of astronomical twilight the pillar is almost invisible, especially above the Moon. The cross increases in intensity (but not necessarily visibility, see later) as twilight advances. The considerable pillar brightening results from the increased reflectivity of the near horizontal ice crystal "mirrors" as the lunar rays angle of incidence approaches 90°. The lesser paraselenae brightening is from reduced scattering losses.

The actual cross visibility worsens as twilight progresses because the increase in halo intensity is more than offset by the tenfold reduction in contrast as the sky brightness increases towards civil twilight. An approximate halo cross visibility window from shortly before the onset of nautical twilight to approaching civil twilight (02:00 to 03:30) is indicated. The time slot supports observation in a monastic environment by monks attending prayer, as few others would be awake.

We now examine variants of cruciform displays. The 806 apparition is portrayed in different ways, as shown in Figure 3. Dots each side of the cross in ASCS (Figure 3) might possibly indicate paraselenae. Other versions have no dots. There is, with one exception, no indication of an upper tangent arc. None show a 22° halo. ASCS resembles a cross *pattée*, perhaps made by tall paraselenae plus upper and lower tangent arcs (Greenler and Mallman, 1972). Figure 6 explores some display variations produced by altering crystal habits, their optical quality and populations.

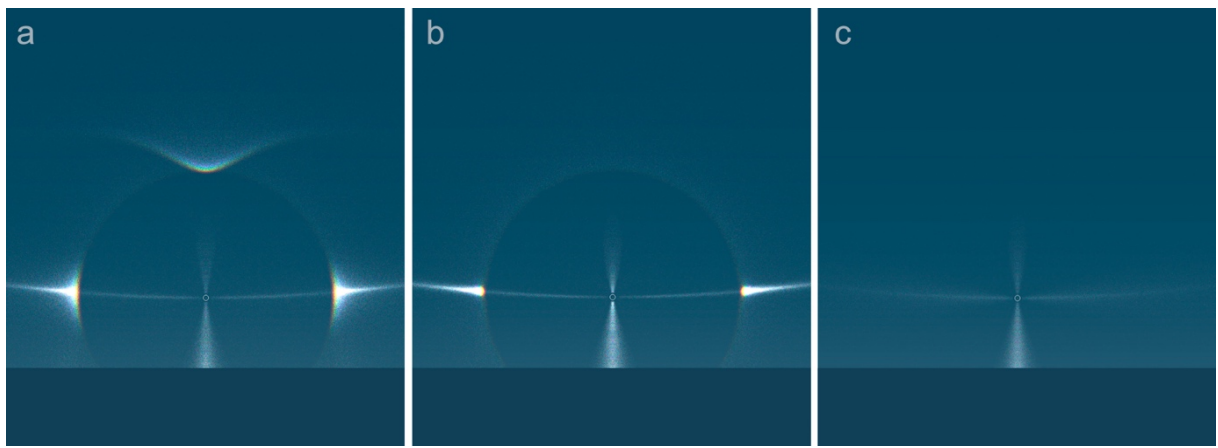

Figure 6: Cross variations for lunar altitude 12.5°: (a) Pillar forming plates with side faces of good optical quality and forming tall paraselenae; (b) As Figure 6(b) but column crystals absent and no tangent arc formed; (c) Crystals as in 4(a) but with internal occlusions or depressed but planar basal faces that block internal rays. A pure cross results with no other halos.

A truncated cross *pattée* (Figure 6a) is readily produced by modifying the high tilt plate crystals that form the pillar. Most pillars come from large imperfect crystals that reflect well from their basal faces but do not otherwise form good halos – the predictions of Figure 5 employ these. Poor crystal quality might be a reason for the scarcity of halo crosses as they simultaneously require a tall pillar *and* a bright paraselenic circle that is only generated by high quality plates or, in rarer circumstances, columns. In Figure 6a the pillar crystals are given better optical quality and they produce prominent extended paraselenae.

Figure 6b shows a display with no upper tangent arc, achieved by removing column crystals from the population.

Figure 6c demonstrates that a lone cross with no other halos is possible. The prediction used only large tilt plate crystals. They were like those of 4(a) *except* that they did not transmit rays internally. Crystals often contain occlusions that interfere with internal ray passage. Some habits also have depressed basal faces that similarly inhibit internal ray paths (see Tape and Moilanen, 2006, pp. 10-19, especially Figure 2.4, 2.5). These crystals thus only form halos by external reflection – the cross of a lunar pillar and paraselenic circle. We are unaware of any modern sighting of this kind.

A 22° halo deliberately appears in our ray tracings. It is there to illustrate the angular scale and is easily eliminated by removing the small concentration of randomly oriented column crystals. However, it is rarely absent in actual displays.

To summarise, lunar halos with the required cruciform aspect can occur but the cross is usually fainter and therefore rarer than surrounding halos. A key issue is whether a cruciform display would be visible during twilight. Quantitative estimates of sky brightness combined with halo predictions by numerical ray tracing indicate a visibility window at latitudes similar to that of Sens of 02:00 to 03:30 local solar time. That in turn supports observation in a monastic environment. Further numerical modelling demonstrates that physically credible variations of ice crystal properties can alter the cross appearance to correspond with the several manuscript renditions. An interesting variant is a pure cross with no other halos. The surprising absence in the records of a tangent arc or a 22° halo, with only one hint at paraselenae, indicates a rare display configuration. Alternatively, these latter halos were present but the scribe ignored them because they were commonplace or because the cross was by far the most important aspect. Regrettably, the manuscripts provide insufficient detail to resolve these issues but what is present is fully consistent with a lunar halo interpretation.

## 4) Possible social impact

Interestingly, these cross-signs in 776 CE and 806 CE are seemingly synchronized with coins with cross-signs in Anglo-Saxon England (see Naismith, 2017) and the Carolingian Empire (Garipzanov, 2008). Sometimes, they manifest remarkable resemblance to the 806 drawing without any clear connection (Figure 7(f), 7(g)). Indeed, it is the very entry of 806 CE in ASCS that Garipsanov (2008, p. 216) cited as evidence of how crosses occupied the minds of clergy of the period, who were dominant in the imperial court, and coins with cross-design introduced soon after this in 818 CE by an imperial edict (Figure 7(h)). Coins can reflect the reality of contemporary society, like the Agnus-Dei coinage in Anglo-Saxon England (Figure 7(a); Keynes and Naismith, 2011); Caesar's comet (C/-43 K1; so-called *sidus Iulium* 'Julius's comet') (Suetonius, 1998; Pliny, 1991; Hisa, 2015; Kronk, 1999) was manifested in contemporary coins in the Roman Empire (Figure 7(b), 7(c); see also Woods, 2012). Also, crosses and celestial objects have already been associated in coins (Figure 7(d), 7(e); Gannon, 2003, p. 74). Therefore, these celestial signs may have cast significant impressions on the contemporary moneyers and die-cutters and prompted them to mint coins with crosses. Alternatively, the increased interest in crosses and symbols prompted increased recording of supportive sky phenomena. Further investigations on the historical celestial events will enhance our understanding not only of the astronomical and meteorological phenomena but also of the human impacts as well (*c.f.*, Silverman, 1998; Odenwald, 2007).

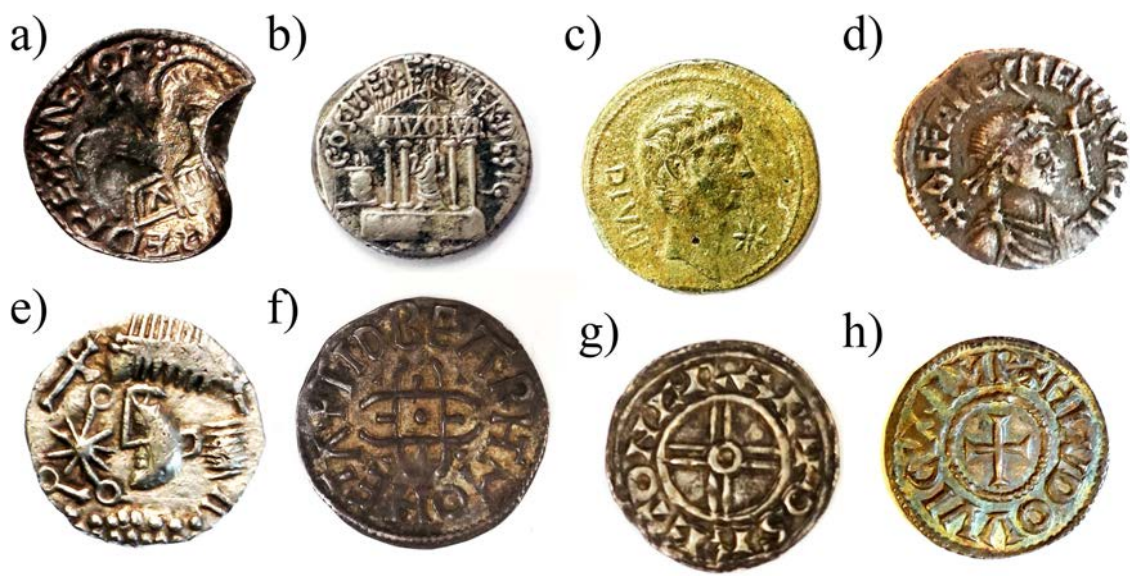

Figure 7: Roman, Anglo-Saxon and Frankish coins with sign of the cross, stars, etc. (courtesy of the Fitzwilliam Museum in the University of Cambridge).

## 5) Conclusion

The philological relationship between six manuscripts containing the 806 lunar record has been examined and their genealogical tree generated; ASCS (the exemplar of AL) is probably the closest to the lost original record (or the copy of it) which was based on the original observation or hearsay at the monastery in Sens in the early 9th century, which also produced AF, while ASMT and ALESVMB compiled in Laon in the mid-9th century represent another tradition. The record in the ASC MS F, compiled in the early 12th century, derives from a lost Winchester Chronicle based on the continental manuscripts, although the exact course of transmission from the continental to the British manuscripts seems distant and unclear. Accordingly, we conclude that the lunar drawing is based on an original observation not in Anglo-Saxon England but one in Champagne, subsequently copied to various chronicles with slight transformations, and was finally imported to Anglo-Saxon England. This is notable, as most of the local celestial events in chronicles have been frequently described without observational site, unless it occurred somewhere outside of the chroniclers' hometown or nearby (*e.g.*, Stephenson, 1997).

Sky brightness calculations and the numerical prediction of possible halos have established a possible window for visibility of a cruciform halo display between 02:00 and 03:30 local solar time on 806 CE June 4 at latitudes similar to that of Champagne (*e.g.*, Sens). This supports observation in a monastic environment. Lunar halo displays of cruciform aspect can occur but the cross is usually fainter and therefore less often seen than surrounding halos. Physically credible variations of ice crystal properties produce crosses corresponding to those of the several manuscript renditions. An interesting variant is a pure cross with no other halos. The surprising absence in the records of the halos that usually accompany the cross indicates a somewhat rare display configuration but then the more commonplace was less likely to be considered worthy of note. Regrettably the manuscripts provide insufficient detail to resolve these issues but what detail is present is fully consistent with a lunar-halo interpretation.

This study also provides an example of the copying process of historical astronomical and meteorological drawings. The record of this lunar event was copied from chronicle to chronicle and spread from Continental Europe to Anglo-Saxon England. As a result of this process, the graphical image was each time slightly modified, as is also seen in the variants of East Asian sunspot drawings. Therefore, when consulting historical astronomical/meteorological drawings, we need to trace their provenance back as far as possible, to best reconstruct the original event.

## Acknowledgement

YU thanks the Fitzwilliam Museum, University of Cambridge for permission to publish the photographs of their coin collection and Rory Naismith for reading the draft and commenting on the coins. LC thanks Leszek Sulich for permission to publish the photograph of a cruciform lunar halo display. We wish to thank the British Library, Biblioteca Apostolica Vaticana, Corpus Christi College of Oxford, Bibliothèque nationale de France, British Library, Stadtbibliothek/Stadtarchiv Trier, and Staatsbibliothek zu Berlin for their permissions to research and reproduce their manuscripts. This research is part of results of JSPS Kakenhi JP18H01254 and JP17J06954.

## References

Baker, P. S., (Ed.): *The Anglo-Saxon Chronicle 8 MS F: a semi-diplomatic edition with introduction and indices*, Cambridge University Press, Cambridge, UK, 2000.

Beard, D.: *Journal of the British Astronomical Association*, 115, 5, 261, 2005.

Boyer, C.B., *The rainbow: From Myth to Mathematics, Princeton*, New Jersey, 1987.

Büntgen, U., Wacker, L., Galván, J. D., *et al.*: 2018, Tree rings reveal globally coherent signature of cosmogenic radiocarbon events in 774 and 993 CE, *Nature Communications*, 9, 3605. DOI: 10.1038/s41467-018-06036-0

Carrasco, V. M. S., Trigo, R. M., and Vaquero, J. M.: Unusual rainbows as auroral candidates: Another point of view, *Publications of the Astronomical Society of Japan*, 69, L1. doi:

10.1093/pasj/psw127, 2017.

Colgrave, B., Mynors, R. A. B. (Ed. and Trans.): *Bede's Ecclesiastical History of the English People*, Oxford Universitye Press, Oxford, UK, 1969.

Cowley, L. T., Schroeder, M: *The HaloSim ray tracing program*, available at https://www.atoptics.co.uk/halo/halfeat.htm, 1998.

Cowley, L. T.: *The Vädersolstavlan, 'Sundog painting'*, available at https://www.atoptics.co.uk/fz346.htm, 2009.

Cubbin, G. P. (Ed.): *The Anglo-Saxon Chronicle 6 MS D: A Semi-diplomatic Edition with Introduction and Indices*, Cambridge University Press, Cambridge, UK, 1996.

Eastwood, B. S.: Early-Medieval Cosmology, Astronomy, and Mathematics, in: *The Cambridge History of Science, vol. 2: Medieval Science*, edited by: Lindberg, D. C., and Shank, M. H., Cambridge University Press, Cambridge, UK, 302-22, 2013.

Egler, R. A.: Astronomical Events Recorded in the Medieval Anglo-Saxon Chronicles, *Journal of the Royal Astronomical Society of Canada*, 96, 5, 184, 2002.

Frobesius, J. N.: *Nova et Antiqua Luminis atque Aurorae Borealis Spectacula*, Helmstadt, 1739.

Fujiyama, M., Hayakawa, H., Iju, T., Kawai, T., Toriumi, S., Otsuji, K., Kondo, K., Watanabe, Y., Nozawa, S. Imada, S.: Revisiting Kunitomo's Sunspot Drawings During 1835 - 1836 in Japan, *Solar Physics*, 294, 43. doi: 10.1007/s11207-019-1429-3, 2019.

Gannon, A.: *The Iconography of Early Anglo-Saxon Coinage: Sixth to Eighth Centuries*, Oxford University Press, Oxford, UK, 2003.

Garipzanov, I. H.: *The Symbolic Language of Authority in the Carolingian World (c. 751–877)*, Brill, Leiden, Netherland, 2008.

Godden, M. R. (Ed. and Trans.): *The Old English History of the World: An Anglo-Saxon Rewriting of Orosius*, Harvard University Press, Cambridge, MA, US, 2016.

Gonzalez, G.: New constraints on ΔT prior to the second century AD, *Monthly Notices of the Royal Astronomical Society*, 482, 1452–1455, doi: 10.1093/mnras/sty2820, 2018.

Greenler, R. G., and Mallmann, A. J.: Circumscribed Halos, *Science*, 176, 4031, 128-131. DOI: 10.1126/science.176.4031.128, 1972.

Härke, H.: Astronomical and Atmospheric Observations in the Anglo-Saxon Chronicle and in Bede, *Antiquarian Astronomer*, 2012, 6, 34-43, 2012.

Hayakawa, H., Iwahashi, K., Ebihara, Y., Tamazawa, H., Shibata, K., Knipp, D. J., Kawamura, A. J., Hattori, K., Mase, K., Nakanishi, I., and Isobe, H.: Long-lasting Extreme Magnetic Storm Activities in 1770 Found in Historical Documents, *The Astrophysical Journal Letters*, 850, L31. doi: 10.3847/2041-8213/aa9661, 2017a.

Hayakawa, H., Iwahashi, K., Tamazawa, H., Toriumi, S., Shibata, K.: Iwahashi Zenbei's Sunspot Drawings in 1793 in Japan, *Solar Physics*, 293, 8. doi:   10.1007/s11207-017-1213-1, 2018.

Hayakawa, H., Mitsuma, Y., Ebihara, Y., Miyake, F.: The Earliest Candidates of Auroral Observations in Assyrian Astrological Reports: Insights on Solar Activity around 660 BCE, *The Astrophysical Journal Letters*, 884, L18. DOI: 10.3847/2041-8213/ab42e4, 2019c

Hayakawa, H., Mitsuma, Y., Fujiwara, Y., Kawamura, A. J., Kataoka, R., Ebihara, Y., Kosaka, S., Iwahashi, K., Tamazawa, H., Isobe, H.: The earliest drawings of datable auroras and a two-tail comet from the Syriac Chronicle of Zūqnīn, *Publ. Astron. Soc. Japan*, 69, 17. doi: 10.1093/pasj/psw128, 2017b.

Hayakawa, H., Stephenson, F. R., Uchikawa, Y., Ebihara, Y., Scott, C. J., Wild, M. N., Wilkinson, J., Willis, D. M.: The Celestial Sign in the Anglo-Saxon Chronicle in the 770s: Insights on Contemporary Solar Activity, *Solar Physics*, 294, 42. doi: 10.1007/s11207-019-1424-8, 2019a.

Hayakawa, H., Willis, D. M., Hattori, K., Notsu, Y., Wild, M. N., and Karoff, C.: Unaided-eye Sunspot Observations in 1769 November: A Comparison of Graphical Records in the East and the West, *Solar Physics*, 294, 95. DOI: 10.1007/s11207-019-1488-5, 2019b.

Hisa, A.: Sidus Iulium and Octavian, *Western History Essays*, Kansai University, 18, 63-76 (in Japanese), 2015.

Irvine, S. (Ed.): *The Anglo-Saxon Chronicle 7 MS E: a semi-diplomatic edition with introduction and indices*, Cambridge University Press, Cambridge, UK, 2004.

Keynes, S., and Naismith, R.: The Agnus Dei pennies of King Æthelred the Unready, *Anglo-Saxon England* 40, 175-223, 2011.

Krisciunas, K., and Schaefer, B. E.: A model of the brightness of moonlight, *Pub. Astr. Soc. Pacific.*, 103, 1033-1039, 1991.

Kronk, G. W.: *Cometography*, 1, Cambridge University Press, Cambridge, UK, 1999.

Lane, A.P., and Irvine, W.M.: Monochromatic phase curves and albedos for the lunar disk, *Astronom. J.*, 78, 267, 1973.

Lewis, S.: *The art of Matthew Paris in the Chronica majora*, Corpus Christi College, Cambridge, 1987

Liebermann, F.: *Ungedruckte anglo-normannische Geschichtsquellen*, Strasbourg, France, 1879.

Martène, E., Durand, U.: *Thesaurus novus anecdotorum*, 3, Paris, France, 1717.

Mekhaldi, F., Muscheler, R., Adolphi, F., Aldahan, A., Beer, J., McConnell, J. R., Possnert, G., Sigl, M., Svensson, A., Synal, H.-A., Welten, K. C., and Woodruff, T. E.: Multiradionuclide evidence for the solar origin of the cosmic-ray events of AD 774/5 and 993/4, *Nature Communications*, 6, 8611. doi: 10.1038/ncomms9611, 2015.

Migne, J.-P. (Ed.): *Ado Viennensis, Usuardus*, *Patrologia Latina*, vol. 123, Paris, France, 1852.

Miyake, F., Nagaya, K., Masuda, K., Nakamura, T.: A signature of cosmic-ray increase in AD 774-775 from tree rings in Japan, *Nature*, 486, 240. doi: 10.1038/nature11123, 2012.

Naismith, R. (Ed.): *Britain and Ireland c. 400–1066, Medieval European Coinage with a Catalogue of the Coins in the Fitzwilliam Museum, Cambridge*, 8, Cambridge University Press, Cambridge, UK, 2017.

Neuhäuser, D. L., and Neuhäuser, R.: "A red cross appeared in the sky" and other celestial signs: Presumable European aurorae in the mid AD 770s were halo displays, *Astronomische Nachrichten*, 336, 913, 2015.

Newton, R. R.: *Medieval Chronicles and the Rotation of the Earth*, John Hopkins University Press, Baltimore, US and London, UK, 1972.

Odenwald, S.: Newspaper reporting of space weather: End of a golden age, Space Weather, 5, 11, S11005, 2007.

Pertz, G. H. (Ed.): *Monnumenta Germaniae Historica, SS II Impensis Bibliopolii Hahniani*, Hannover, Germany, 1829.

Pertz, G. H. (Ed.): *Monumenta Germaniae Historica, SS I, Impensis Bibliopolii Hahniani*, Hannover, Germany, 1826.

Pertz, G. H. (Ed.): *Monumenta Germaniae Historica, SS IV Impensis Bibliopolii Hahniani*, Hannover, Germany, 1841.

Pertz, G. H. (Ed.): *Monumenta Germaniae Historica, SS XV Impensis Bibliopolii Hahniani*, Hannover, Germany, 1888.

Pevtsov, A., Griffin, E., Grindlay, J., Kafka, S., Bartlett, J., Usoskin, I., Mursula, K., Gibson, S., Pillet, V., Burkepile, J., Webb, D., Clette, F., Hesser, J., Stetson, P., Munoz-Jaramillo, A., Hill, F., Bogart, R., Osborn, W., and Longcope, D.: Historical astronomical data: urgent need for preservation, digitization enabling scientific exploration, *Bulletin of the American Astronomical Society*, 51, 190, 2019.

Pliny the Elder: *Naturalis Historia, Natural history. Vol.1 : Praefatio; Libri I-II*, revised edition, The Loeb classical library, 330, edited and translated by: Rackham, H., Harverd University Press, Cambridge, MA, US and London, UK, 1991.

Schlegel, B., Schlegel, K.: *Polarlichter zwischen Wunder und Wirklichkeit*, Spektrum Akademischer Verlag, Heidelberg, 2011.

Schröer, N.: *Die Annales S. Amandi Und Ihre Verwandten: Untersuchungen zu einer Gruppe karolingischer Annalen des 8. Und frühen 9. Jahrhunderts*, Göppingen, Germany, 1975.

Silverman, S. M.: Early auroral observations, *J. Atmos. Sol.-Terr. Phys.*, 60, 10, 997. doi: 10.1016/S1364-6826(98)00040-6, 1998.

Smith, N.: The Crab nebula and the class of Type IIn-P supernovae caused by sub-energetic electron-capture explosions, *Monthly Notices of the Royal Astronomical Society*, 434, 102-113. doi: 10.1093/mnras/stt1004, 2013.

Sôma, M., and Tanikawa, K.: Earth rotation derived from occultation records, *Publications of the*

*Astronomical Society of Japan*, 68, 29. doi: 10.1093/pasj/psw020, 2016.

Stephenson, F. R.: *Historical Eclipses and Earth's Rotation*, Cambridge, Cambridge University Press, 1997.

Stephenson, F. R., and Green, D. A.: *Historical supernovae and their remnants*, Clarendon Press, Oxford, UK, 2002.

Stephenson, F. R., Morrison, L. V., and Hohenkerk, C. Y.: Measurement of the Earth's rotation: 720 BC to AD 2015, *Proceedings of the Royal Society A*, 472, 2196, 20160404, 2016.

Stephenson, F. R., Willis, D. M.: The earliest drawing of sunspots, *Astronomy & Geophysics*, **40**, 6.21-6.22, 1999.

Stephenson, F. R., Willis, D. M., Hayakawa, H., Ebihara, Y., Scott, C. J., Wilkinson, J., and Wild, M. N.: Do the Chinese Astronomical Records Dated AD 776 January 12/13 Describe an Auroral Display or a Lunar Halo? A Critical Re-examination, *Solar Physics*, 294, 4, 36. doi: 10.1007/s11207-019-1425-7, 2019.

Stephenson, F. R., Yau, K. C. C., Hunger, H.: Records of Halley's comet Babylonian tablet, *Nature*, 314, 587-592, 1985.

Stephenson, F. R.: Astronomical evidence relating to the observed 14C increases in A.D. 774-5 and 993-4 as determined from tree rings, *Adv. Space Res.*, 55, 1537. doi: 10.1016/j.asr.2014.12.014, 2015.

Suetonius: *De Vita Caesarum, Lives of the Caesars Books I-IV, revised edition*, The Loeb classical library, 31, edited and translated by: Rolfe, J. C., Harvard University Press, Cambridge, MA, US and London, UK, 1998.

Swanton, M. J.: *The Anglo-Saxon Chronicles*, Phoenix, London, UK, 2000.

Tape, W., and Moilanen, J.: *Atmospheric Halos and the Search for Angle X*, American Geophysical Union, Washington DC, US, 2006.

Tape, W.: *Atmospheric Halos*, American Geophysical Union, Washington DC, US, 1994.

Usoskin, I. G., Kovaltsov, G. A., Mishina, L. N., Sokoloff, D. D., and Vaquero, J.: An Optical Atmospheric Phenomenon Observed in 1670 over the City of Astrakhan Was Not a Mid-Latitude Aurora, *Solar Physics*, 292, 15. doi: 10.1007/s11207-016-1035-6, 2017.

Usoskin, I. G., Kromer, B., Ludlow, F., Beer, J., Friedrich, M., Kovaltsov, G. A., Solanki, S. K., and Wacker, L.: The AD775 cosmic event revisited: the Sun is to blame, *Astron. Astrophys.*, 552, L3. doi: 10.1051/0004-6361/201321080, 2013.

Usoskin, I. G.: A history of solar activity over millennia, *Living Reviews in Solar Physics*, 14, 3. doi: 10.1007/s41116-017-0006-9, 2017.

Uusitalo, J., Arppe, L., Hackman, T., *et al.*: 2018, Solar superstorm of AD 774 recorded subannually by Arctic tree rings, *Nature Communications*, 9, 3495 DOI: 10.1038/s41467-018-05883-1

Vaquero, J. M., and Vazquez, M.: *The Sun Recorded Through History: Scientific Data Extracted*

*from Historical Documents*, Springer, Berlin, Germany, 2009.

Vaquero, J. M.: Halo with unusual parhelion, *Weather*, 57, 82, 2002.

Vaquero, J. M.: Historical sunspot observations: A review, *Advances in Space Research*, 40, 929-941. doi: 10.1016/j.asr.2007.01.087, 2007.

Vidier, A.: *L'historiographie à Saint-Benoît-sur-Loire et les Miracles de Saint Benoît*, Paris, France, 1965.

Willis, D. M., Stephenson, F. R.: Solar and auroral evidence for an intense recurrent geomagnetic storm during December in AD 1128, *Annales Geophysicae*, 19, 289-302, 2001.

Weiss, P.: The vision of Constantine, *Journal of Roman Archaeology*, 16, 237-259, 2003.

Woods, D.: Postumus and the Three Suns: Neglected Numismatic Evidence for a Solar Halo, *The Numismatic Chronicle*, 172, 85-92, 2012.

Appendix: The texts and the translations of the 806 entries in the relevant Manuscripts

Texts

ASCS (Perz, 1826, p. 103): *806. Anno incarnationis dominicae 806. pridie Non. Iunii, luna 14. signum crucis mirabili modo in luna apparuit feria 5. prima aurora incipiente, quasi hoc modo ✦. Eodem anno 3. Kal. Septembris luna 12. die dominica hora quarta, corona mirabilis in circuitu solis apparuit.*

ASMT (Pertz, 1841, p. 6): *806. prid. Non. Iun. luna 14. signum crucis mirabili modo in luna apparuit hoc modo # feria 5. prima aurora incipiente. Eodem anno 3. Kal. Sept. luna 12. die dominica, hora 4. corona mirabilis in circuitu solis apparuit.*

ALESVMB (Pertz, 1888, p. 1294): *806. pridie Non. Iun., luna [14, si]gnum crucis mirabili [modo in] luna apparuit hoc [modo] # feria 5ta, prima [aurora] incipiente. Eodem anno [3. Kal.] Septembris, luna 12a, die [domini]co, hora 4ta, corona mira[bilis in c]ircuitu solis apparuit.*

AL (Martène and Durand, 1717, col. 1401) : *Anno Incarnationis dominicæ DCCCVI. II. nonas Junii, luna XIV. signum crucis mirabili modo in luna apparuit feria V. prima aurora incipiente, quasi hoc modo #. Eodem anno, III. cal. Septembris, luna XII. die dominica, hora IV. corona mirabilis in circuitu solis apparuit.*

AF (Vidier, 1965, p. 218): *DCCCVI Anno incarnationis dominicae dcccvi II Non. iun. luna.XIIII. signum crucis mirabili modo in luna apparuit, feria .V. prima aurora incipiente quasi hoc modo.*

ASC MS F (Baker, 2000, p. 59): (Old English Version): *Eac on ðys ylcan geare,.ii. nonas Iunii, rodetacn wearð ateowe[d] on ðan monan anes Wodnesdæges innan ðare dagenge. An eft on ðis geare an kalendas Septembris an wunderlic tre[nd]el wearð ateowed abutan ðare sunnan. . . .*

(Latin Version): *Hoc anno etiam, .ii. nonas Iunii, luna quarta decima, signum crucis mirabili modo in luna apparuit feria .v. aurora incipiente, hoc modo #. Eodem anno, .iii. kalendas Septembris, luna xii., die dominica, hora. .iiii., corona mirabilis in circuitu solis apparuit.*

A 12th-century chronicle founded on WC (London, British Library, Cotton Nero A. viii, f. 25r-25v; Liebermann, 1879, pp. 63-64): *Hoc anno, 2 non. Junii, luna 14, signum crucis mirabili modo in luna apparuit, feria quinta, prima aurora incipiente, quasi hoc modo #. Eodem anno 3 kal. Septembris, luna 12, die dominica, hora 4, corona mirabilis in circuitu solis apparuit.*

Translations

ASCS: 806. In the 806th year from the incarnation of the Lord, on 4th June, the 14th day of the Moon, the sign of the cross in a remarkable fashion, appeared at the Moon on Thursday at the first dawn, like this #. In the same year, on 30 August, the 12th day of the Moon, on Sunday, at the fourth hour, a wonderful crown appeared around the Sun.

ASMT: 806. On 4[th] June, the 14[th] day of the Moon, the sign of the cross in a remarkable fashion, appeared at the Moon, like this #, on Thursday at the first dawn. In the same year, on 30 August, the 12[th] day of the Moon, on Sunday, at the fourth hour, a [wonderful] crown appeared [around] the Sun.

ALESVMB: 806. On 4th June, [the 14th day] of the Moon, [the sign] of the cross in a remarkable [fashion], appeared [at] the Moon, like this #, on Thursday at the first [dawn]. In the same year, [on 30 August], the 12th day of the Moon, [on Sunday], at the fourth hour, a wonderful crown appeared around the Sun.

AL: In the 806th year from the incarnation of the Lord, on 4th June, the 14th day of the Moon, the sign of the cross in a remarkable fashion, appeared at the Moon on Thursday at the first dawn, like this #. In the same year, on 30 August, the 12th day of the Moon, on Sunday, at the fourth hour, a wonderful crown appeared around the Sun.

AF: In the 806th year from the incarnation of the Lord, on 4th June, the 14th day of the Moon, the sign of the cross in a remarkable fashion, appeared at the Moon on Thursday at the first dawn, like this #.

ASC MS F: (Old English Version): Also in this same year, on 4 June, the sign of the cross appeared at the moon one Wednesday at the dawning: and again in this year, on 30 August, an amazing ring appeared around the sun.

(Latin Version): Also in this year, on 4th June, the 14th day of the Moon, the sign of the cross in a remarkable fashion, appeared at the Moon on Thursday at dawn, like this #. In the same year, on 30 August, the 12th day of the Moon, on Sunday, at the fourth hour, a wonderful crown appeared around the Sun.

A 12th-century chronicle founded on WC: In this year, on 4th June, the 14th day of the Moon, the sign of the cross in a remarkable fashion, appeared at the Moon on Thursday at the first dawn, like this #. In the same year, on 30 August, the 12th day of the Moon, on Sunday, at the fourth hour, a wonderful crown appeared around the Sun.